\documentstyle[multicol,aps,prl,epsf]{revtex}
\begin{document}
\title{Large-Deviation Functions for Nonlinear Functionals of a \\
Gaussian Stationary Markov Process} 

\author{Satya N. Majumdar$^{(1),(2)}$ and Alan J. Bray$^{(3)}$}

\address{(1)Laboratoire de Physique Quantique (UMR C5626 du CNRS),
Universit\'e Paul Sabatier, 31062 Toulouse Cedex, France. \\
(2) Tata Institute of Fundamental Research, Homi Bhabha Road,
Mumbai-400005, India. \\
(3)Department of Physics and Astronomy, University of Manchester,
Manchester, M13 9PL, UK}

\date{\today}

\maketitle

\begin{abstract}

\noindent  We introduce  a general  method,  based on  a mapping  onto
quantum  mechanics,  for  investigating  the large-$T$  limit  of  the
distribution   $P(r,T)$   of   the   nonlinear  functional   $r[V]   =
(1/T)\int_0^T dT'\,V[X(T')]$, where $V(X)$ is an arbitrary function of
the stationary Gaussian  Markov process $X(T)$. For $T  \to \infty$ at
fixed  $r$   we  obtain   $P(r,T)  \sim  \exp[-\theta(r)   T]$,  where
$\theta(r)$ is a large-deviation function. We present explicit results
for a number of special  cases including $V(X) = XH(X)$ (where  $H(X)$ 
is the Heaviside function), which is related to the  cooling  and  the 
heating degree days relevant to weather derivatives.

\end{abstract}

\begin{multicols}{2}

\section{Introduction}
The ``persistence''  of a continuous  stochastic process has  been the
subject of  considerable recent interest among  both theoreticians and
experimentalists   in   the   field   of   nonequilibrium   processes.
Persistence  is  the probability  $P(t)$  that  a stochastic  variable
$x(t)$ of  zero mean  does not  change sign up  to time  $t$.  Systems
studied include reaction-diffusion processes, phase-ordering kinetics,
fluctuating  interfaces  and  simple  diffusion  from  random  initial
conditions \cite{review}.  Experimental measurements have been carried
out  on breath  figures \cite{marcos},  liquid  crystals \cite{yurke},
soap  froths \cite{tam},  and diffusion  of  Xe gas  in one  dimension
\cite{wong}.   In  ``coarsening'' systems  like  these,  which do  not
possess  a  definite length  or  time  scale,  the persistence  has  a
power-law  decay, $P(t)  \sim  t^{-\theta}$, at  late  times, and  the
persistence  exponent  $\theta$ is  in  general  non-trivial. In  such
systems, the  normalized two-time correlation  function, $C(t_1,t_2) =
\langle  x(t_1)x(t_2)   \rangle  /[\langle  x^2(t_1)  \rangle\,\langle
x^2(t_2)  \rangle]^{1/2}$  has  the  ``scaling'' form,  $C(t_1,t_2)  =
f(t_1/t_2)$, depending  only on the ratio  of the two  times.  In such
systems, a  simplification is achieved by  introducing the logarithmic
time-scale $T=\ln  t$, and the  normalized variable $X(T)=x(t)/\langle
x^2(t)\rangle^{1/2}$,  since the correlation  function of  $X$ depends
only on  the time  difference $T_2-T_1$, i.e.\  the process  $X(T)$ is
stationary.   Thus  one  is  led  to  consider  stationary  stochastic
processes. These processes  are, of course, also of  interest in their
own  right.   In  the  new  time  variable,   the  persistence  decays
exponentially, $P \sim \exp(-\theta T)$.

In this paper we consider the integrated quantity
\begin{equation}
r = \frac{1}{T} \int_0^T dT'\,V[X(T')]
\label{rdef}
\end{equation}
where  $V(X)$ is  an  arbitrary function  of  the stochastic  variable
$X(T)$.   Hence $r$  is  a functional  of  $V$. For  the special  case
$V(X)=H(X)$, where $H(X)$ is the  Heaviside step function, $r$ is just
the fraction of  time for which $X(T')>0$ in the  time interval $O \le
T' \le  T$. In this  case, the probability distribution,  $P(r,T)$, of
$r$ for given $T$ is  just the ``residence -time'' distribution which,
together with  the related  ``sign-time'' distribution, where  $V(X) =
{\rm  sgn}(X)$,  has attracted  a  lot  of  attention earlier  in  the
mathematics  literature\cite{Kac,motd}   and  more  recently   in  the
statistical physics  community\cite{std}.  Here we  consider a general
function $V(X)$.  We restrict our  attention, however, to the class of
processes where  $X(T)$ is a  stationary Gaussian Markov  process, for
which some  analytic progress can be  made.  A case  of some interest,
which we analyze in detail, is $V(X)=X H(X)$. This  case is  relevant  
to  weather derivatives as we now explain.

It has been estimated that about  1 trillion dollars of the 7 trillion
dollar  US economy  is  weather related  \cite{weather}. For  example,
weather conditions directly affect agricultural outputs and the demand
for   energy  products   and  indirectly   affect   retail  businesses
\cite{Cao}.   Weather  derivatives,  first  introduced  in  1997,  are
financial  instruments which  allow hedging  (by, for  example, energy
suppliers) against  adverse weather conditions over a  period of time.
Here ``adverse'' might mean an  unusually warm winter, when low demand
for  energy  would  affect  the  supplier's profits,  as  well  as  an
unusually cold  one when  the supplier is  unable to meet  the demand.
Temperature derivatives, the most  common form of weather derivatives,
are based  on the  concepts of ``heating  degree days''  and ``cooling
degree days'', which are (rough) measures of the cumulative demand for
heating and cooling respectively.

Let $X(T)$ be the temperature at time  $T$ in a given city. On a given
day, $n$, the  mean, $X_n$, of the highest  and lowest temperatures is
recorded.  The number of cooling degree days ($CDD$), over a period of
$N$  days, is  given by  $CDD =  \sum_{n=1}^N  \max(X_n-X_o,0)$, where
$X_o$ is  a reference, or  baseline, temperature, while the  number of
heating degree  days ($HDD$) is $HDD  = \sum_{n=1}^N \max(X_o-X_n,0)$.
In the  present paper  we will, for  simplicity, use an  integral over
continuous time rather than a sum, so that the cooling degree days are
given  by $CDD =  \int_0^T dT'\,  [X(T')-X_o] \,H[X(T')  - X_o]$,
where $H(X)$  is the Heaviside  step function. Thus $CDD$  is the
integrated  temperature   excess  (over  the   reference  temperature)
restricted  to  those  periods  where  the temperature  is  above  the
reference level.  It is a crude  measure of the amount of cooling (air
conditioning) required  during the period  $T$ and also of  the energy
required to produce  this cooling. Note that the  power consumption of
an air conditioner actually varies, for small temperature differences,
as  the square  of the  temperature  difference between  the room  and
ambient temperatures, so a better measure of the energy required would
be  $\int_0^T dT'\,[X(T')-X_o]^2 \,H[X(T') -  X_o]$, instead of $CDD$. 
In the  notation of this paper, $CDD = rT$,  with $V(X) = (X-X_o)
H(X-X_o)$  in Eq.\  (\ref{rdef}).  The  number of  heating degree
days   ($HDD$)   is  similarly   given   by   $HDD   =  \int_0^T   dT'
[X_o-X(T')]\,H[X_o-X(T')]$,  which  is a  measure  the amount  of
heating  required in  the period  $T$ and  of the  energy  required to
produce this  heating.  Estimating the likelihood  of large deviations
from  the mean  in $CDD$  or $HDD$  is clearly  of interest  to energy
companies.

To make further progress, a realistic statistical model of temperature
fluctuations is required. At  this stage, however, any realistic model
is  too intractable to  allow analytical  progress. To  illustrate the
general approach we will  instead employ a simple, though unrealistic,
model in  which the temperature, $X(T)$,  is taken to  be a stationary
Gaussian  Markov process.   We will  discuss the  limitations  of this
model,  and possible improvements,  in the  Conclusion.  As  a further
simplification, we take the reference temperature $X_o$ to be the mean
of  $X(T)$, though  this simplification  is not  essential and  can be
relaxed.

The outline of the paper is as follows. In section II we introduce the
general  approach  to  the  problem  of  computing  the  distribution,
$P_T(r)$, of the  quantity $r$ defined by Eq.\  (\ref{rdef}).  Using a
path-integral  representation, the calculation  of $P_T(r)$  is mapped
onto  a problem  in quantum  mechanics, in  which the  function $V(X)$
appears in the  potential energy. In the limit of  large $T$, only the
quantum ground  state energy is  required. The final result  takes the
form $P_T(r)  \sim \exp[-\theta(r)T]$, where  the function $\theta(r)$
is  a  large-deviation  function.   For  the  case  of  the  sign-time
distribution, corresponding  to $V(X)={\rm  sgn}(X)$, $r$ lies  in the
range $-1 \le r  \le 1$, and it is clear that  $\theta(1)$ is just the
usual persistence exponent, $\theta$, since $r=1$ requires $X(T') > 0$
for $0 \le T'  \le T$. In section III, the method  is illustrated on a
number  of special  cases, of  which $V(X)=\alpha  X$  and $V(X)=\beta
X^2/2$  are studied  first as  exactly-soluble  tutorial illustrations
before turning to the CDD  problem (in the simple form outlined above)
and finally the  sign-time distribution. The last two  examples can be
solved    analytically   in    various   regimes,    and   numerically
elsewhere. Section IV  concludes with a discussion and  summary of the
results.

\section{GENERAL APPROACH}
\label{general}
Consider    the   general    stationary   Gaussian    Markov   process
(Ornstein-Uhlenbeck process) $dx/dt = -\mu x + \xi(t)$, where $\xi(t)$
is Gaussian white noise with zero mean, and correlator $\langle \xi(t)
\xi(t') \rangle = 2D\,\delta(t-t')$.  After the change of variables $t
= T/\mu$, $x=(2D/\mu)^{1/2}X$, the equation takes the form
\begin{equation}
\frac{dX}{dT} = - X + \eta(T)\ ,
\label{ou}
\end{equation}
where $\langle \eta(T) \rangle = 0$ and
\begin{equation}
\langle \eta(T) \eta(T') \rangle = \delta(T-T')\ .
\label{noisecorr}
\end{equation}

The probability distribution of $X(T')$ for  $0 \le T' \le T$ is given
by
\begin{equation}
P[X]  = N  \exp\left\{-\frac{1}{2}\int_0^T  dT' \left[\dot{X}^2(T')  +
X^2(T') \right]\right\}\ ,
\label{prob}
\end{equation}
where $\dot{X}(T')\equiv dX/dT'$, and $N$ is a normalization constant.
Our goal is to calculate  the probability distribution of $r$, defined
by equation (\ref{rdef}). In practice  it is convenient to look at the
distribution $P_u(u)$ of the quantity $u=rT$. Its Laplace transform is
\begin{equation}
\tilde{P}_u(s) = \langle \exp(-rsT) \rangle = Z(s)/Z(0)\ ,
\end{equation}
where $Z(s)$ is given by the path integral
\begin{eqnarray}
Z(s) &  = & \int DX(T)\exp\{-\frac{1}{2}\int_0^T dT'  (\dot{X}^2 + X^2
\nonumber \\ && \hspace{4.5cm} + 2s\,V[X])\}\ .
\end{eqnarray}

We are  interested in the  limit $T \to  \infty$. It is  convenient to
impose   periodic  boundary   conditions,   $X(T)=X(0)$,  since   this
restriction  will  not change  the  results  in  the large-$T$  limit.
Furthermore, the  exponential in (\ref{prob})  should strictly contain
the combination $(\dot{X} + X)^2$ instead of $(\dot{X}^2 + X^2)$.  The
missing  term, $2X\dot{X}$,  is  however a  perfect derivative,  whose
integral  vanishes for  periodic boundary  conditions.   Finally, with
these boundary  conditions the  function $Z(s)$ is  the imaginary-time
Feynman path-integral  that gives the partition function  of a quantum
particle  with Hamiltonian  ${\cal  H} =  p^2/2  + X^2/2  + sV(X)$  at
inverse temperature $T$, $p$ being the canonical momentum conjugate to
$X$. For  $T \to \infty$ the  ground state dominates to  give, in this
limit,
\begin{equation}
\langle \exp(-rsT) \rangle = \exp\{-T[E(s)-E(0)]\}\ ,
\label{asymptotic}
\end{equation}
where  $E(s)$  is  the  ground-state energy  for  the  Schr\"{o}dinger
equation
\begin{equation}
-\frac{1}{2} \frac{d^2\psi}{dX^2} + U(X)\psi = E(s)\psi\ ,
\label{TISE}
\end{equation}
with potential
\begin{equation}
U(X) = X^2/2 + sV(X)\ .
\label{potential}
\end{equation}
For $s=0$  the problem  reduces to a  simple harmonic  oscillator, and
$E(0)=1/2$.

The  stochastic  process  $x(t)$  studied  above  corresponds  to  the
position  of  a  Brownian  particle  in  an  external  potential  $\mu
x^2/2$. For the case of  a pure Brownian motion ($\mu=0$), Kac derived
a  formalism\cite{Kac}   to  study  the   distributions  of  arbitrary
functionals $V[x]$  which also used  a mapping to  the Schr\"{o}dinger
equation.   Note, however,  that the  method presented  above  for the
$\mu\ne 0$ case differs in details from the original Kac formalism.

To  illustrate  the method  we  discuss  two  simple examples,  before
turning to some nontrivial cases, including the CDD problem.

\section{SPECIAL CASES}
\label{special}
\subsection{$V(X) = \alpha X$}
For this case we have
\begin{equation}
u =rT =\alpha\int_0^T dT'\,X(T')\ .
\end{equation}
This  case is actually  trivial since  $u$, being  a sum  of zero-mean
Gaussian variables,  is itself a zero-mean Gaussian  variable.  All we
require, therefore, is the variance, given by
\begin{equation}
\langle u^2  \rangle = \alpha^2 \int_0^T dT_1  \int_0^T dT_2\, \langle
X(T_1)\,X(T_2) \rangle\ .
\label{variance}
\end{equation}
For the  Ornstein-Uhlenbeck process (\ref{ou}),  with noise correlator
(\ref{noisecorr}), one easily  finds $\langle X(T_1)\,X(T_2) \rangle =
(1/2)\exp(-|T_1  -T_2|)$. Inserting  this result  in (\ref{variance}),
and  extracting the  leading  large-$T$ behavior,  gives $\langle  u^2
\rangle  \to  \alpha^2 T$,  and  therefore  $\langle  r^2 \rangle  \to
\alpha^2/T$.   Hence the asymptotic  distribution of  $r$ is  given by
(neglecting prefactors)
\begin{equation}
P_T(r) \sim \exp(-Tr^2/2\alpha^2)\ ,\ \ \ \ \ T \to \infty\ .
\label{example1}
\end{equation}

We  now show  how the  general  machinery we  have set  up in  section
\ref{general}  recovers  this result.   The  potential  $U(X)$ in  the
Schr\"odinger equation (\ref{TISE}) takes the form
\begin{equation}
U(x) = X^2/2 + s\alpha X = (X + s\alpha)^2/2 - s^2\alpha^2/2\ .
\end{equation}
This is just  a harmonic oscillator with a shifted  origin, so $E(s) =
1/2 - s^2\alpha^2/2$ and, using (\ref{asymptotic}),
\begin{equation}
\langle \exp(-rsT) \rangle = \exp(Ts^2\alpha^2/2)
\end{equation}
for large $T$. To recover $P_T(r)$ we can invert the Laplace transform
as follows. Neglecting pre-exponential factors,
\begin{equation}
P_T(r) =  T \int_{-i\infty}^{i\infty} \frac{ds}{2\pi  i}\, \exp[T(rs +
s^2\alpha^2/2)]\ .
\end{equation}
This integral can, of course,  be evaluated exactly. Here, however, we
use the method of steepest descents,  which is valid for large $T$ and
can be readily  generalized to the other, less  trivial, cases that we
will  discuss.  Writing the  integrand in  the form  $\exp[Tg(s)]$, we
have $g(s)  = rs  + s^2\alpha^2/2$. The  integral is dominated  by the
saddle point at  $s = -r/\alpha^2$, where $g  = - r^2/2\alpha^2$.  The
integration contour is  deformed to pass over the  saddle point, which
lies on  the real  $s$ axis.  The  saddle point  is thus a  minimum of
$g(s)$  with respect to  variations of  $s$ along  the real  axis. The
final  result, ignoring  non-exponential prefactors,  is  identical to
(\ref{example1}).

The structure of (\ref{example1}) is
\begin{equation}
P_T(r) \sim \exp[-\theta(r)T]\ ,
\label{extremes}
\end{equation}
with
\begin{equation}
\theta(r) = r^2/2\alpha^2\ .
\label{theta_X}
\end{equation}

We  can  easily  generalize  this  method to  arbitrary  $V(X)$.   The
path-integral   approach   gives   the  general   result   (neglecting
prefactors)
\begin{equation}
P_T(r) \sim \int_{-i\infty}^{i\infty} ds\,\exp[Tg(s)]\ ,
\end{equation}
where
\begin{equation}
g(s) = rs + E(0) - E(s)\ .
\end{equation}
Using the steepest-descent method for the integral gives
\begin{equation}
\theta(r) = \max_{s}\, \left[E(s) - \frac{1}{2} -rs\right]\ ,
\label{max}
\end{equation}
where  we have  inserted $E(0)=1/2$.   The  next example  is a  simple
application of this idea.

\subsection{$V(X) = \beta X^2/2$}
This case illustrates the power of the method. The potential energy is
now
\begin{equation}
U(X) = (1+\beta s)X^2/2 \ ,
\end{equation}
so we have a harmonic  oscillator again but with a modified frequency,
$\omega = (1+\beta s)^{1/2}$, giving $E(s)=(1+\beta s)^{1/2}/2$. Thus
\begin{eqnarray}
\theta(r)    &   =   &    \max_s\,\left[\frac{1}{2}(1+\beta   s)^{1/2}
-\frac{1}{2}    -     rs    \right]    \nonumber    \\     &    =    &
\left(\sqrt{\frac{r}{\beta}}-
\frac{1}{4}\sqrt{\frac{\beta}{r}}\right)^2\ .
\label{thetar2}
\end{eqnarray}
Note that $\theta(r)$  now has its minimum at $r  = \beta/4$, which is
just the  mean value of $r$  [noting that $\langle X^2  \rangle = 1/2$
follows from  (\ref{ou}) and  (\ref{noisecorr})], while large  ($r \to
\infty$)  and   small  ($r  \to   0$)  values  of  $r$   are  strongly
suppressed. An  expansion of $\theta$  near its minimum value  gives a
Gaussian             distribution             $P_T(r)             \sim
\exp[-4T(r-\beta/4)^2/\beta^2]$. This agrees  with the result expected
from    the    central    limit    theorem,   i.e.\    $P_T(r)    \sim
\exp[-(r-\mu)^2/2\sigma^2]$,  with mean $\mu  = \beta/4$  and variance
$\sigma^2 = \beta^2/8T$. For  $T \to \infty$, the distribution becomes
very narrow such  that, at fixed $r$, the  central limit theorem fails
to give the correct asymptotics.  The form (\ref{extremes}) thus gives
the behavior  in the extreme tails  of the distribution  at large $T$.
The  function  $\theta(r)$  is  a ``large-deviation  function''  which
controls the distribution of $r$ for large $T$.

We  note that  this  special  case with  $V(X)=\beta  X^2/2$ was  also
studied  recently  by  Farago\cite{Farago}  in the  context  of  power
fluctuations  in  the  Langevin  equation  (\ref{ou})  by  a  somewhat
different method.  The probability  density function of the dissipated
power in reference\cite{Farago} is precisely the distribution $P_T(r)$
studied   here   and   the  corresponding   large-deviation   function
$\theta(r)$  has  the  same  expression as  in  Eq.   (\ref{thetar2}).
However our derivation seems simpler and easily generalizable to other
forms of $V(X)$ as we show below.

\subsection{$V(X) = XH(X)$: ``Cooling Degree Days''}
In this case  the quantity $rT$ gives the integrated  value of $X$ over
the interval $T$,  restricted to those values where  $X>0$.  If $X$ is
the excess temperature over  some baseline value where cooling becomes
necessary, $rT$  is a  crude measure of  the total  energy consumption
required to  provide the  cooling [$V(X) =  X^2 H(X)$ would  be a
better  measure, as  discussed in  the Introduction].  In  the present
simple model we have taken the mean temperature equal to the reference
temperature (with  $X(T)$ being the  deviation from the  mean), though
this  is not  an  essential restriction.   The Schr\"odinger  equation
(\ref{TISE}) takes the form
\begin{eqnarray}
\psi'' - X^2\psi -2sX\psi + 2E(s)\psi & = & 0,\ \ \ X \ge 0\ ,
\label{tidepos} \\
\psi'' - X^2\psi + 2E(s)\psi & = & 0,\ \ \ X \le 0\ ,
\label{tideneg}
\end{eqnarray}
where $\psi''  \equiv d^2\psi/dX^2 $.   The required solutions  can be
expressed in  terms of  parabolic cylinder functions,  ${\rm D}_p(x)$,
using the standard solutions of the parabolic cylinder equation $y'' -
(x^2/4 +  a)y=0$ \cite{AS}. Selecting  the solutions that  satisfy the
physical boundary condition $\psi(\pm \infty) = 0$ gives
\begin{eqnarray}
\psi(X) & =  & \psi_+(X) = A\,{\rm D}_{p_+}\left(\sqrt{2}(X+s)\right),
\ \ \ X \ge 0\ ,  \\ & = & \psi_-(X) = B\,{\rm D}_{p_-}\left(-\sqrt{2}
X\right), \ \ \ X \le 0\ ,
\end{eqnarray}
where $A$, $B$ are constants and
\begin{eqnarray}
p_+ & =  & E(s) - \frac{1}{2} +  \frac{s^2}{2}\ , \\ p_- & =  & E(s) -
\frac{1}{2}\ .
\end{eqnarray}
The ratio  $A/B$ and the  energy eigenvalues $E(s)$ are  obtained from
matching  the  wavefunction and  its  derivative  at  $X=0$, i.e.\  we
require    $\psi_+(0)    =     \psi_-(0)$    and    ${\psi_+}'(0)    =
{\psi_-}'(0)$. This yields the eigenvalue equation
\begin{equation}
\frac{{{\rm   D}_{p_+}}'(\sqrt{2}s)}{{\rm  D}_{p_+}(\sqrt{2}s)}   =  -
\frac{{{\rm D}_{p_-}}'(0)}{{\rm D}_{p_-}(0)}\ .
\label{eigenvalue}
\end{equation}
The  determination of  $E(s)$  from these  equations  is not  possible
analytically for general $s$. However,  in the regimes $s \to \infty$,
and  $s \to  -\infty$,  which determine  the  small-$r$ and  large-$r$
behavior respectively of $\theta(r)$, analytical progress is possible.
We consider these in turn.

\subsubsection{The limit $r \to \infty$}
As we shall  see, in this limit it is sufficient  to compute $E(s)$ in
the  limit $s \to  -\infty$.  This  can either  be done  directly from
(\ref{eigenvalue}),   or  using   the  following   (simpler)  physical
argument.

Recall  that, for  $V(X) =  XH(X)$, the  potential $U(X)$  in the
Schr\"odinger equation is
\begin{eqnarray}
U(X) & = & X^2/2 + sX\ ,\ \ \  X \ge 0\ , \nonumber \\ & = & X^2/2\ ,\
     \ \ X \le 0\ .
\end{eqnarray}
For  $s \to  -\infty$,  the potential  has  a deep  minimum, of  depth
$-s^2/2$, located  at $X=-s$. The wavefunction  is exponentially small
at $X=0$, $\psi(0) \sim \exp(-s^2/2)$, and the change in the potential
in the  regime $X<0$ has an  exponentially small effect  on the ground
state energy. Thus
\begin{equation}
E(s) = -s^2/2 + 1/2 + O(e^{-s^2}), \ \ \ s\to -\infty\ .
\end{equation}
The same  result can  be obtained, after  some algebra,  directly from
(\ref{eigenvalue}).   Inserting   the  result  in   (\ref{max})  gives
$\theta(r) = \max_s(-s^2/2 - rs) = r^2/2$. Since the maximum occurs at
$s = -r$, the calculation is self-consistent for $r \to \infty$.  Thus
we obtain
\begin{equation}
\theta(r) \to r^2/2\ ,\ \ \ \ r \to \infty\ .
\label{max_cdd_larger}
\end{equation}

This large-$r$  result has the same form  as equation (\ref{theta_X}),
which gives  the general-$r$  result for potential  $V(X) =  \alpha X$
(with $\alpha=1$ here). This  correspondence is not so surprising: For
$r \to  \infty$, the dominant  processes $X(T')$, for $0\le  T'\le T$,
will correspond  to large  positive $X$, so  the step function $H(X)$ 
in $V(X)$ plays no role in this limit.

\subsubsection{The limit $r \to 0$}

We shall see that this limit corresponds to $s \to \infty$. Again, one
can use physical arguments as a short cut. For $s \to \infty$, one has
$U(X) = X^2/2$ for $X<0$, with  essentially a hard wall at $X=0$. This
gives, to leading order, $E(s) \to 3/2$. We need, however, the leading
correction to this result. Since  the wave function does not penetrate
far into the wall, we can  neglect the $X^2$ term in the potential for
$X >  0$, i.e.\  write $U(X) =  sX$.  The Schr\"odinger  equation then
simplifies to
\begin{eqnarray}
\psi'' - 2sX\,\psi + 2E(s)\psi & = & 0\  ,\ \ \ X \ge 0\ , \\ \psi'' -
X^2\psi + 2E(s)\psi & = & 0\ ,\ \ \ X \le 0\ .
\end{eqnarray}
The wavefunction for $X \le 0$ is again a parabolic cylinder function,
while for $X \ge 0$ it can be expressed as an Airy function:
\begin{eqnarray}
\psi(X) &  = & A\,{\rm Ai}\left((2s)^{1/3}X\right)\  , \ \ \  X \ge 0\
,\\ & = & B\,{\rm D}_{p_-}\!\left(-\sqrt{2}X\right)\  , \ \ \ X \le 0\
,
\end{eqnarray}
where  $A$, $B$  are constants,  and  $p_- =  E(s) -  1/2$ as  before.
Matching the wavefunction and its  first derivative at $X=0$ gives the
eigenvalue equation
\begin{equation}
\frac{(2s)^{1/3}{\rm  Ai}'(0)}{\sqrt{2}{\rm  Ai}(0)}  = -  \frac{{{\rm
 D}_{p_-}}'(0)}{{\rm       D}_{p_-}(0)}      =      \frac{\sqrt{2}{\rm
 \Gamma}\left(\frac{1-p_-}{2}\right)}                             {{\rm
 \Gamma}\left(-\frac{p_-}{2}\right)}\ ,
\label{matching}
\end{equation}
where ${\rm  \Gamma}(x)$ is the Gamma  function.  In the  limit $s \to
\infty$, the left-hand side  of (\ref{matching}) tends to infinity, so
the right-hand side (RHS) must also diverge in this limit.  The ground
state  corresponds  to  the  first  divergence,  where  $p_-  \to  1$.
Therefore we write $p_- =  1 - \epsilon$ in (\ref{matching}), and seek
the leading  behavior as $\epsilon \to  0$. This gives  ${\rm RHS} \to
-(2/\pi)^{1/2}/\epsilon$.   Inserting this result  in (\ref{matching})
gives, to leading order,
\begin{eqnarray}
\epsilon   &   =  &   -   2^{1/6}  \sqrt{\frac{2}{\pi}}\,   \frac{{\rm
 Ai}(0)}{{\rm    Ai}'(0)}\,   s^{-1/3}    \nonumber   \\    &    =   &
 \sqrt{\frac{2}{\pi}}\,\frac{2^{1/6}}{3^{1/3}}\,             \frac{{\rm
 \Gamma}(1/3)}{{\rm \Gamma}(2/3)}\, s^{-1/3} \equiv a\,s^{-1/3}\ ,
\end{eqnarray}
the last equation  defining the constant $a$. Finally  we have $E(s) =
p_- + 1/2 = 3/2 -  \epsilon$. Inserting this in (\ref{max}) gives, for
$r \to 0$,
\begin{equation}
\theta(r) = \max_s(1 - as^{-1/3} - rs) = 1 - br^{1/4}\ ,
\label{max_cdd_smallr}
\end{equation}
where
\begin{equation}
b  = 4\left(\frac{a}{3}\right)^{3/4}  = \frac{4\sqrt{2}}{3\pi^{3/8}}\,
\left(\frac{{\rm   \Gamma}(1/3)}{{\rm   \Gamma}(2/3)}\right)^{3/4}   =
2.04763\ldots
\end{equation}
Note  that  the   value  of  $s$  at  which   the  maximum  occurs  in
(\ref{max_cdd_smallr}) is $s =  (a/3r)^{3/4}$, justifying our use of a
large-$s$ analysis of $E(s)$ for the limit $r \to 0$.

That $\theta(r) \to 1$ for $r \to 0$ is intuitively clear, since $r=0$
requires  $X(T)  \le  0$  for  all  $T$. This  reduces  to  the  usual
persistence probability  of the  Markov process (\ref{ou}),  for which
$\theta=1$.

\subsubsection{$\theta(r)$ for $r$ near $\langle r \rangle$}
Equations   (\ref{max_cdd_larger})  and   (\ref{max_cdd_smallr})  give
analytical results  for $\theta(r)$ in  the limits of large  and small
$r$  respectively.   For general  $r$,  $\theta$  has  to be  computed
numerically.   There is,  however, one  other regime  where analytical
progress is  possible, namely for $r$  close to its  mean value, where
the central limit theorem (CLT) applies.

The mean value is given by
\begin{eqnarray}
\langle  r \rangle &  = &  \langle V(X)  \rangle =  \langle X H(X)
\rangle\ \nonumber \\ & = & \frac{1}{2}(\langle X \rangle +\langle |X|
\rangle)\ .
\end{eqnarray}
The Gaussian distribution for $X$ gives immediately $\langle X \rangle
= 0$ and $\langle |X| \rangle = \sqrt{2/\pi} \langle X^2 \rangle^{1/2}
= 1/\sqrt{\pi}$. Thus
\begin{equation}
\langle r \rangle = \frac{1}{2\sqrt{\pi}}\ .
\end{equation}
In a similar way, the variance of $r$ can be calculated, by exploiting
the Gaussian properties  of the process $X(T)$. A tedious but
straightforward calculation gives, for $T \to \infty$,
\begin{equation}
\sigma^2  \equiv  \langle  r^2  \rangle  -  \langle   r  \rangle^2  =
\frac{1}{2\pi T}\,\left(\pi + \ln 2 - 2\right)\ .
\label{var}
\end{equation}
The central limit theorem then  gives the behavior of $P_T(r)$ for $r$
near  $r_c$  as  $P_T(r)  = (2\pi\sigma^2)^{-1/2}  \exp[-(r-\langle  r
\rangle)^2/2\sigma^2]$.    Inserting  the  explicit   expressions  for
$\langle r \rangle$ and  $\sigma^2$ gives $P_T(r) \sim \exp[-\theta(r)
T]$, with
\begin{equation}
\theta(r)  = \left(\frac{\pi}{\pi  +  \ln 2  -  2}\right)\, \left(r  -
\frac{1}{2\sqrt{\pi}}\right)^2\ ,
\end{equation}
correct to leading nontrivial order in $(r -\langle r \rangle)$.

\begin{figure}
\narrowtext\centerline{\epsfxsize\columnwidth
\epsfbox{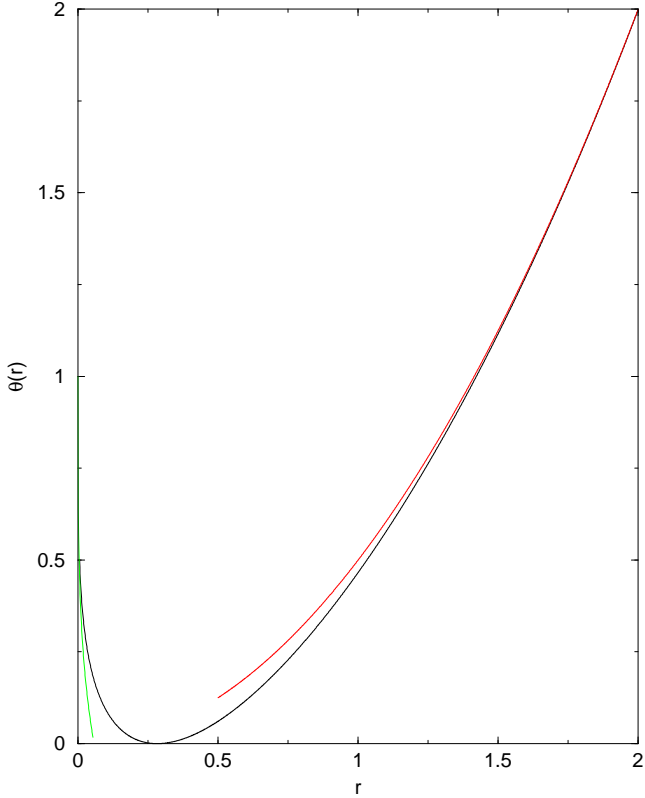}}
\caption{The function $\theta(r)$ for the CDD problem, showing the
asymptotic behavior for small and large $r$.}
\label{f1}
\end{figure}

The full result for $\theta(r)$ can be obtained by numerically solving
(\ref{eigenvalue}) for  the ground state energy $E(s)$  for each value
of $r$, then using  (\ref{max}) to find the corresponding $\theta(r)$.
The result is displayed in  Figure \ref{f1}, with the asymptotic forms
for $r \to  \infty$ and $r \to 0$ indicated. Note  the very sharp rise
to   the  value   unity  as   $r  \to   0$,  as   indicated   by  Eq.\
(\ref{max_cdd_smallr}).

In terms of the CDD problem, the behavior below the minimum (i.e.\ for
$r <  \langle r \rangle$)  determines the probability of  an unusually
small  $CDD$ (``cool  summer''), while  the region  above  the minimum
corresponds to  an unusually large  $CDD$ (``hot summer'').   The fact
that the function $\theta(r)$  initially increases less rapidly to the
right of the minimum than to the left indicates that (within this very
simple model)  summers with a  slightly larger than average  $CDD$ are
more probable than  those with a slightly smaller  than average $CDD$.
This  asymmetry is  a consequence  of the  nonlinear  relation between
$CDD$ and the temperature  fluctuations, which are symmetric about the
mean in our  model. It should be stressed  that the integration period
$T$  has been  taken  to be  large,  to justify  the steepest  descent
calculation. In practice this means  that $T$ (the length of a summer,
say) must be large compared to the correlation time of the temperature
(a few days, perhaps), which is not totally unreasonable.

\subsection{$V(X) = {\rm sgn}(X)$: The ``Sign-Time Distribution''}
The structure of  the calculation is similar to  that of the preceding
subsection. The Schr\"odinger equation is
\begin{eqnarray}
\psi'' - X^2\psi  - 2 [s - E(s)]\psi &  = & 0,\ \ \  X>0\ ,\\
\psi'' - X^2\psi  + 2 [s + E(s)]\psi & = & 0,\ \ \ X< 0\ .
\end{eqnarray}
The solutions are parabolic cylinder functions,
\begin{eqnarray}
\psi(X) & = & \psi_+(X) =  A\,{\rm D}_{p+}(\sqrt{2}X)\ , \ \ \ X\ge 0\
,\\ & = & \psi_-(X) = B\,{\rm D}_{p_-}(-\sqrt{2}X)\ ,\ \ \ X\le 0\ .
\end{eqnarray}
where now
\begin{equation}
p_{\pm} = E(s) \mp s - 1/2\ .
\label{p_std}
\end{equation}
Matching  the  wavefunction and  its  derivative  at  $X=0$ gives  the
eigenvalue equation
\begin{equation}
\frac{{{\rm   D}_{p_+}}'(0)}{{\rm   D}_{p_+}(0)}   =   -   \frac{{{\rm
D}_{p_-}}'(0)}{{\rm D}_{p_-}(0)}\ ,
\label{eigenvalue_std}
\end{equation}
which, using  standard identities  relating ${\rm D}_p(0)$  and ${{\rm
D}_p}'(0)$ to gamma functions \cite{AS}, reduces to
\begin{equation}
\frac{{\rm                         \Gamma}\left(\frac{1-p_+}{2}\right)}
{{\rm\Gamma}\left(-\frac{p_+}{2}\right)}      =      -      \frac{{\rm
\Gamma}\left(\frac{1-p_-}{2}\right)}
{{\rm\Gamma}\left(-\frac{p_-}{2}\right)}\ .
\label{gamma_std}
\end{equation}
Although this equation cannot  be solved analytically for general $s$,
the limits $s \to 0$ and $s \to \infty$ are tractable. Note that $E(s)
= E(-s)$ by symmetry, so there  is no need to consider $s \to -\infty$
separately.

The analysis starts  from the potential well, $U(X)  = X^2/2 + s\,{\rm
sgn}(X)$.   For   small  $s$,  the  ground  state   energy  $E(s)$  is
perturbatively    close    to,     but    slightly    smaller    than,
$E(0)=1/2$. Therefore we write
\begin{equation}
E(s) = 1/2 - \delta(s)\ ,
\end{equation}
where   we  anticipate   $\delta(s)  =   O(s^2)$  from   the  symmetry
$E(s)=E(-s)$. Inserting this  form in (\ref{p_std}), (\ref{gamma_std})
becomes
\begin{equation}
\frac{{\rm                    \Gamma}\left(\frac{1+s+\delta}{2}\right)}
{{\rm\Gamma}\left(\frac{s+\delta}{2}\right)}     =     -    \frac{{\rm
\Gamma}\left(\frac{1-s+\delta}{2}\right)}
{{\rm\Gamma}\left(\frac{-s+\delta}{2}\right)}\ .
\end{equation}
Expanding to fourth order in $s$ and second order in $\delta$ gives
\begin{equation}
\delta = s^2\,\ln 2 - cs^4 + O(s^6)\ ,
\end{equation}
where
\begin{equation}
c  =  \frac{1}{3}(\ln 2)^3  +  \frac{\pi^2}{12}\ln2 -  \frac{1}{4}{\rm
\zeta}(3) = 0.269865\ldots\ ,
\end{equation}
and  ${\rm \zeta}(n)$ is  the Riemann  zeta function.   Inserting this
result in (\ref{max}) gives
\begin{eqnarray}
\theta(r) & = & \max_s[-rs - s^2\ln  2 + cs^4 + O(s^6)] \nonumber \\ &
 = & \frac{r^2}{4\ln 2} + c\left(\frac{r}{2\ln 2}\right)^4 + O(r^6)\ .
\end{eqnarray}
The maximum  occurs at $s = -r/2\ln2  + O(r^3)$, so our  study at small
$s$ is self-consistent at small $r$.

\begin{figure}
\narrowtext\centerline{\epsfxsize\columnwidth
\epsfbox{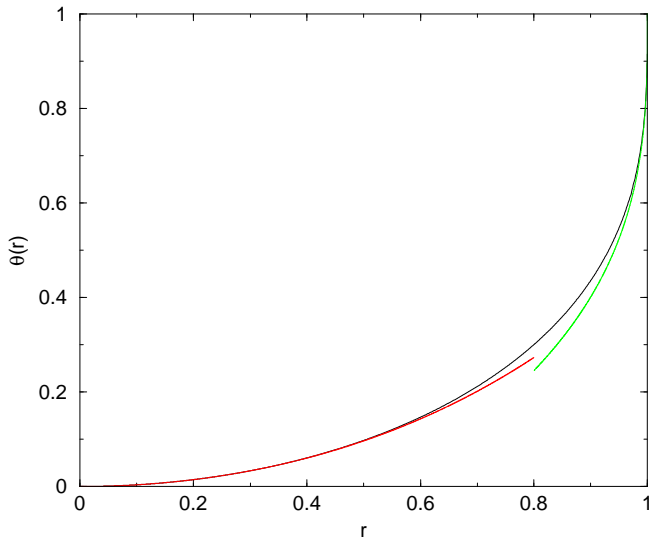}}
\caption{The function $\theta(r)$ for the sign-time distribution,
showing the limiting behavior for $r \to 0$ and $r \to 1$.}
\label{f2}
\end{figure}

For $s \to  \infty$, on the other hand, the  potential develops a hard
wall  at the  origin,  and  has a  depth  of $-s$  next  to the  wall.
Therefore  we write  $E(s) =  -s +  3/2 -  \epsilon$,  with $\epsilon$
small. Putting this form in (\ref{p_std}), (\ref{gamma_std}) becomes
\begin{equation}
\frac{{\rm  \Gamma}[s+\epsilon/2]}{{\rm\Gamma}[s+(\epsilon-1)/2]}  = -
\frac{{\rm \Gamma}[\epsilon/2]}{{\rm\Gamma}[(\epsilon-1)/2]}\ .
\end{equation}
Taking the limits $s \gg 1$, and $\epsilon \ll 1$ readily leads to
\begin{equation}
\epsilon = (\pi s)^{-1/2}
\end{equation}
to leading order for large $s$. Using this in (\ref{max}) gives
\begin{eqnarray}
\theta(r) & = & \max_s[1 -  (1+r)s - 1/(\pi s)^{1/2}] \nonumber \\ & =
 & 1 - \frac{3}{2}\left(\frac{2}{\pi}(1+r)\right)^{1/3} + \cdots
\end{eqnarray}
The maximum  occurs at $s=[2\sqrt{\pi}(1+r)]^{-2/3}]$,  which tends to
infinity as $r  \to -1$ so the calculation  is self-consistent in this
limit. The symmetry of the problem  under $r \to -r$ leads to the more
general result
\begin{equation}
\theta(r)  = 1 -  \frac{3}{2}\left(\frac{2}{\pi}(1-|r|)\right)^{1/3} +
\cdots\ ,\ \ \ \ r \to \pm 1\ .
\end{equation}
The function $\theta(r)$ is plotted  in Figure \ref{f2}, with only the
region $r \ge 0$ shown explicitly. The limiting behavior for small $r$
and $r$ close to unity is also shown.

\section{Conclusion}
In this  paper we  have presented a  general method for  computing the
asymptotic  behavior  of  the  the  probability  distribution  of  the
quantity  $r  =  (1/T)\int_0^T  dT'\,V[X(T')]$,  where  $V(X)$  is  an
arbitrary  function  and $X(T)$  is  an Ornstein-Uhlenbeck  stochastic
process representing the motion of a Brownian particle in the presence
of a  stable parabolic potential $\mu  X^2/2$. The main  new result is
that for $\mu>0$,  the distribution of $r$, for  large window size $T$
has the form $P_T(r) \sim \exp[-\theta(r)T]$, where $\theta(r)$ is the
large-deviation function.  The calculation proceeds via a mapping onto
a quantum  mechanical problem described by  the Schr\"odinger equation
(\ref{TISE}) for a particle moving in the potential (\ref{potential}),
where  $s$ is  the Laplace  variable  conjugate to  $rT$. The  inverse
Laplace  transform can  be performed  using steepest  descents  in the
limit $T \to \infty$.

Two  non-trivial applications  have been  presented.  The  first  is a
calculation  of  the  large-deviation  function  $\theta(r)$  for  the
cooling degree  days problem, based  on a simple model  of temperature
fluctuations. The model used  assumes that the temperature fluctuation
$X(T)$  is  described by  the  Ornstein-Uhlenbeck process  (\ref{ou}),
i.e.\  that  $X(T)$  is  a  stationary Gaussian  Markov  process  with
time-independent noise strength. This may not be a realistic model for
several reasons: (i)  A simple Markov process is not  thought to be an
optimized  model of  temperature fluctuations,  which tend  to exhibit
stronger  autocorrelations than  a Markov  process.  A  more plausible
statistical  model, according to  \cite{Cao}, writes  the fluctuation,
$X_n$, of the mean (average of  daily high and low) temperature on day
$n$  as the  linear  combination $X_n  =  \sum_{m=1}^k w_{n-m}\,X_m  +
\eta_n$,  where  the  ``memory  kernel''  $w_{n-m}$  is  a  decreasing
function  of $n-m$ and  $\eta_n$ is  uncorrelated Gaussian  noise. The
Markov case corresponds to $k=1$. Here $X_n$ is the difference between
the measured mean  temperature on day $n$ and  its expected value. The
latter  should   contain  a   365  day  seasonal   variation  (roughly
sinusoidal)  (ii) The  noise strength  should also  contain a  365 day
seasonal variation:  the variance of the  temperature fluctuations can
be  different at  different times  of the  year.  (iii)  The reference
temperature for the  calculation of $CDD$ and $HDD$  should in general
be different  from the expected  temperature.  We hope  to incorporate
some of these features in future studies.

The   second   non-trivial  application   is   to  the   ``sign-time''
distribution.  In the  context of ``sign-time", we point  out that the
asymptotic  form  of  the  ``sign-time" distribution  has  a  markedly
different behavior in the Ornstein-Uhlenbeck process (where a Brownian
particle moves  in a stable  parabolic potential $\mu  X^2/2$) studied
here compared to the ordinary  Brownian motion ($\mu=0$). In the later
case,  the ``sign-time"  distribution $P_T(r)$  is independent  of the
window  size  $T$   for  all  $T$  and  is   given  by  $P_T(r)=1/{\pi
\sqrt{1-r^2}}$\cite{motd}. In the former  case ($\mu>0$), on the other
hand, the  ``sign-time" distribution depends explicitly  on the window
size $T$  and $P_T(r)\sim \exp[-\theta(r)T]$  for large $T$  where the
large-deviation function $\theta(r)$ has been computed exactly in this
paper.

We  further note  that  for this  ``sign-time"  problem, the  function
$\theta(r)$  can also  be  obtained using  the ``independent  interval
approximation'' (IIA)  \cite{IIA1,IIA2}, which exploits  the fact that
the intervals between zero crossings are statistically independent for
a  Markov process.   In fact,  for  renewal type  processes where  the
successive intervals  are statistically independent,  the ``sign-time"
distribution has  been computed by Godr\`eche  and Luck\cite{GL} using
the interval size distribution as an input. Their result can be simply
extended   to  calculate   the  ``sign-time'   distribution   for  the
Ornstein-Uhlenbeck process.  The  IIA also has the virtue  that it can
be  used  to  obtain  approximate results  for  non-Markov  processes.
Persistence exponents, for example,  are often given rather accurately
by the IIA \cite{review}. However,  it is not straightforward to adapt
this  IIA method to  general nonlinear  functions $V(X)$,  whereas the
path-integral approach and mapping onto quantum mechanics adopted here
is readily applicable to any $V(X)$.

Here we  have only considered Gaussian Markov  processes mainly because
they are simple and  amenable to analytical calculations. Recently the
calculations of the asymptotic  distributions for the ``sign-time" and
other related  quantities such as ``local-time" have  been extended to
non-Gaussian Markov  processes where a  Brownian particle moves  in an
arbitrary stable or unstable potential and moreover exact results have
been obtained\cite{MC}  when the underlying potential is  random as in
the Sinai model.  The extension of these methods and results presented
here to  non-Markov processes,  however, still remains  as one  of the
outstanding challenges for the future.

\section{Acknowledgment}
SM thanks A. Comtet for useful discussions.

\end{multicols}

\end{document}